\documentclass[twocolumn,showpacs,preprintnumbers,amsmath,amssymb]{revtex4}

\usepackage{graphicx}
\usepackage{dcolumn}
\usepackage{bm}

\begin{document}

\preprint{APS/PRL}

\title{Static magnetic order in metallic K$_{0.49}$CoO$_{2}$}

\author{J. Sugiyama$^1$}
 \email{e0589@mosk.tytlabs.co.jp}
\author{H. Nozaki$^1$}
\author{Y. Ikedo$^1$}
\author{K. Mukai$^1$}
\author{J. H. Brewer$^2$}
\author{E. J. Ansaldo$^3$}
\author{G. D. Morris$^3$}
\author{D. Andreica$^{4,5}$}
\author{A. Amato$^5$}
\author{T. Fujii,$^6$}
\author{A. Asamitsu$^6$}
\affiliation{%
$^1$Toyota Central Research and Development Labs. Inc., 
 Nagakute, Aichi 480-1192, Japan}%

\affiliation{$^2$TRIUMF, CIAR and 
Department of Physics and Astronomy, University of British Columbia, 
Vancouver, BC, V6T 1Z1 Canada 
}%

\affiliation{$^3$TRIUMF, 4004 Wesbrook Mall, Vancouver, BC, V6T 2A3 Canada 
}%

\affiliation{$^4$Faculty of Physics, Babes-Bolyai University, 3400 Cluj-Napoca, Romania
}%

\affiliation{$^5$Laboratory for Muon-Spin Spectroscopy, 
Paul Scherrer Institut, Villigen PSI, Switzerland 
}%

\affiliation{$^6$Cryogenic Center, University of Tokyo, 
2-11-16 Yayoi, Bunkyo-ku, Tokyo 113-0032, Japan
}%

\date{\today}

\begin{abstract}
By means of muon spin spectroscopy, 
we have found that K$_{0.49}$CoO$_2$ crystals undergo successive magnetic transitions 
from a high-$T$ paramagnetic state to a magnetic ordered state below 60~K 
and then to a second ordered state below 16~K, 
even though K$_{0.49}$CoO$_2$ is metallic at least down to 4~K. 
An isotropic magnetic behavior and wide internal-field distributions suggest 
the formation of a commensurate helical spin density wave ({\sf SDW}) state below 16~K, 
while a linear {\sf SDW} state is likely to exist above 16~K. 
It was also found that K$_{0.49}$CoO$_2$ exhibits a further transition at 150~K 
presumably due to a change in the spin state of the Co ions. 
Since the $T$ dependence of the internal-field below 60~K 
was similar to that for Na$_{0.5}$CoO$_2$, 
this suggests that magnetic order is more strongly affected
by the Co valence 
than by the interlayer distance/interaction and/or the charge-ordering.
\end{abstract}


\maketitle
The Na$_x$CoO$_2$ (NCO) system has emerged recently 
as a new paradigm of low-dimensional, 
strongly correlated systems 
because of its direct relevance to several condensed matter topics of great
current interest, 
primarily enhanced thermopowers,\cite{NCO_3} 
frustrated magnetic ordering in triangular lattices,\cite{muSR_6}  
and unconventional superconductivity;\cite{NCO_sc_1}
and the possible fundamental interrelationship between those properties,
which can be studied as a function of doping as the rich but complex
NCO phase diagram is explored further.\cite{muSR_5,Foo_1}
In particular, magnetic and superconducting order parameters are
based on the CoO$_2$ planes, as for the cuprates, 
with interlayers playing a crucial role for the doping and couplings. 
In the NCO case mobility and inhomogeneity of the Na$^+$ ions
may lead to extra difficulties in the interpretation of
experimental results on magnetic ordering, especially in the
metallic region for $x\geq$~0.7. 
In addition, and in contrast to the cuprates,
the spin-charge dynamics of the Co ions in the plane play a major role
which needs to be understood better, by further studies.

Since NCO is a member of the family 
$A_x$CoO$_2$ ($A$=Li, Na, K, Rb and Cs), and other members 
also display interesting and 
complex magnetic orderings depending on the average Co valence of the CoO$_2$ planes 
($V_{\rm Co}$), we have initiated its systematic study 
with positive muon spin rotation and relaxation
($\mu^+$SR) spectroscopy. 
Results for $A$=Li and $x$=1 were reported recently,\cite{muSR_8}
and in this Letter we wish to report the appearance of
successive ordering transitions to quasi-static 
(i.e., static at least for muon's life time, 2.2~$\mu$s)
bulk magnetic states in K$_{0.49}$CoO$_2$, and in contrast lack of magnetic
order above 5~K for Rb$_{0.3}$CoO$_2$.
Both K$_{0.49}$CoO$_2$ and Rb$_{0.3}$CoO$_2$ crystals 
exhibit metallic conductivity down to 4~K 
with nearly $T$ independent susceptibility ($\chi$), 
while the $\chi(T)$ curve of the two crystals showed 
a small increase below $\sim$20~K
for reasons unknown.\cite{KCO_1} 
It is thus believed that both compounds are 
Pauli paramagnets down to 4~K.

For the cobaltites with $V_{\rm Co}\leq3.4$, 
we have previously proposed a universal dome-shaped magnetic phase diagram not only for 
NCO but also for more complex layered cobaltites.\cite{muSR_6,muSR_5} 
Very recently, the dome-shape was reconfirmed by the $\mu^+$SR experiments on 
Na$_{0.85}$CoO$_2$ and NaCoO$_2$.\cite{Mendels_1,Mendels_2}, 
although $T_{\rm N}$ for Na$_{0.85}$CoO$_2$ seems to be a little higher than 
that predicted by a smooth dome function, perhaps due to structural changes over
a narrow range of $x$  for NCO.   
This leads to the question of what are  the common intrinsic features 
for all layered cobalt oxides 
in the $V_{\rm Co}$ range above 3.4. 
In particular, $V_{\rm Co}\sim$3.5 is worth investigating, 
because Na$_{0.5}$CoO$_2$ was reported 
to enter into a charge-ordered insulating state below 83~K and/or 53~K, 
by $\chi$ and resistivity ($\rho$) measurements.\cite{Foo_1} 
Also, $\mu^+$SR on Na$_{0.5}$CoO$_2$ 
showed the appearance of quasi-static magnetic order below 83~K, 
for which several spin-precession signals were detected.\cite{Mendels_1} 
The origin of magnetism was naturally explained 
using the concept of charge-ordering in the CoO$_2$ plane. 


High quality single-crystal platelets of K$_{0.49}$CoO$_2$ were grown 
at the University of Tokyo by a flux technique
using reagent grade K$_2$CO$_3$ 
and Co$_3$O$_4$ powders as starting materials.
A mixture of KCl, K$_2$CO$_3$ and B$_2$O$_3$ 
was used as the flux.
The typical dimension of the obtained K$_{0.49}$CoO$_2$ platelets
was $\sim 3\times3\times$0.05~mm$^3$. 
The composition of the platelets was determined by an induction coupled plasm analysis. 
The preparation and characterization of these crystals 
were reported in greater detail elsewhere.\cite{KCO_1} 
In order to stop muons in the sample and to increase the signal intensity, 
$\sim$30 platelets were stacked in a muon-veto sample holder. 
The $\mu^+$SR experiments were performed on the surface muon beam line 
of {\bf M20} at TRIUMF and {\bf $\pi$M3} (GPS) at PSI.  The direction of the spin
of the muons relative to the plane of the crystals
was set by the spin rotator - Wien filters in the beamlines. 
The experimental setup and procedures is described elsewhere.\cite{SDW_2}

\begin{figure}
\includegraphics[width=6cm]{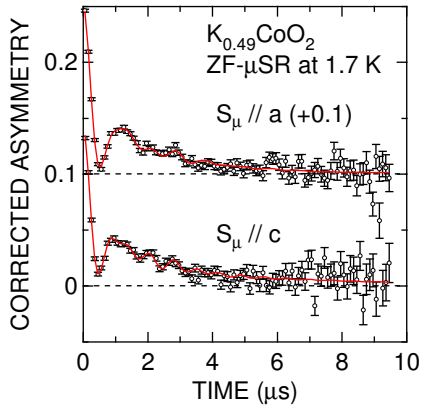}
\caption{\label{fig:ZF-muSR1} ZF-$\mu^+$SR time spectra of 
  single crystal platelets of K$_{0.49}$CoO$_2$ at 1.7~K. 
  The configurations of the sample and the initial muon spin direction 
  ${\bf S}_\mu(0)$ are 
  (top) ${\bf S}_\mu(0) \parallel {\bf a}$ and 
  (bottom) ${\bf S}_\mu(0) \parallel {\bf c}$.  
  The top spectrum is offset by 0.1 for clarity of the display.
}
\end{figure}
Figure~\ref{fig:ZF-muSR1} shows zero-field (ZF-)$\mu^+$SR time spectra at 1.7 K 
for single crystal platelets of K$_{0.49}$CoO$_2$. 
The top spectrum was obtained with the initial $\mu^+$ spin direction 
${\bf S}_\mu(0)$ parallel to the $a$-axis 
and the bottom one with ${\bf S}_\mu(0)$ parallel to $c$. 
A clear oscillation due to quasi-static internal fields ${\bm H}_{\rm int}$ 
is observed in both cases, 
indicating an isotropic magnetic structure in K$_{0.49}$CoO$_2$. 
This behavior is very different from the large magnetic anisotropy observed 
in the other layered cobaltites; such as,  
Na$_{0.9}$CoO$_2$, [Ca$_2$CoO$_3$]$_{0.62}$[CoO$_2$] 
and [Ca$_2$Co$_{4/3}$Cu$_{2/3}$O$_4$]$_{0.62}$[CoO$_2$].\cite{muSR_6,muSR_1}
The signal is fitted best by three exponentially relaxed cosine oscillations 
with the same initial phase, 
which describe three different muon sites 
in a commensurate ${\bm H}_{\rm int}$ distribution. 

\begin{figure}
\includegraphics[width=5.5cm]{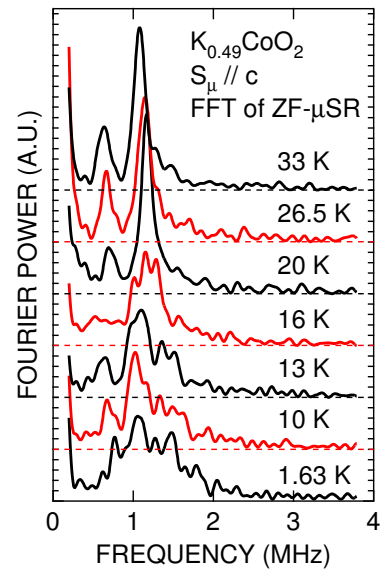}
\caption{\label{fig:FFT} Temperature dependence of the 
  Fourier Transform of the ZF-$\mu^+$SR time spectrum 
  for K$_{0.49}$CoO$_2$. 
  Here, the ZF-$\mu^+$SR spectrum was obtained 
  using ${\bf S}_\mu(0) \parallel {\bf c}$ configuration.  
}
\end{figure}
Figure~\ref{fig:FFT} shows the change in 
the Fourier Transform of 
the ZF-$\mu^+$SR time spectrum as a function of $T$. 
At the lowest $T$ measured, 
the spectrum consists of three main peaks at around 0.8, 1.1 and 1.5~MHz 
with additional small peaks, 
although to the eye the spectrum seems to be a main broad peak 
with several minor peaks, 
which is due to a wide distribution of ${\bm H}_{\rm int}$ in K$_{0.49}$CoO$_2$. 
This reflects the appearance of 
a commensurate spin density wave ({\sf C-SDW}) order, 
as predicted by the calculation using a Mott-Hubbard model, discussed later.

As $T$ increases from 1.63~K, 
the peak at 0.8~MHz shifts towards a lower frequency, 
and then seems to disappear at $\sim$20~K, 
while the other two peaks (at 1.1 and 1.5~MHz) are roughly $T$ independent.   
The spectrum drastically changes its shape at 20~K with further increasing $T$. 
That is, the FFT spectrum consists of two sharp peaks above 20~K, 
showing that the field distribution above 20~K is narrower than below 20~K. 
In other words, this indicates a  change in the magnetic structure 
probably due to a transition around 20~K. 
The frequencies of both peaks decrease with further increasing $T$ up to $\sim$60~K.    

\begin{figure}
\includegraphics[width=7cm]{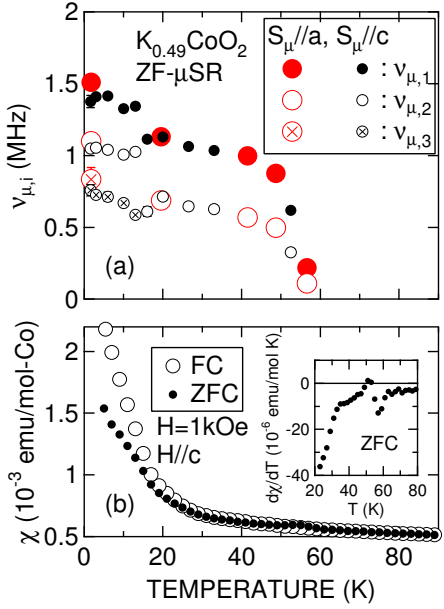}
\caption{\label{fig:ZF-muSR}  Temperature dependence of 
(a) muon precession frequency $\nu_{\mu, i}$ and 
(b) dc susceptibility $\chi$ for K$_{0.49}$CoO$_2$.
The data of $\nu_{\mu, i}$ were obtained by fitting the ZF-spectrum 
with three exponentially relaxed cosine oscillation signals with the same initial phase. 
Large circles represent the data obtained with ${\bf S}_\mu(0) \parallel {\bf a}$ configuration at TRIUMF, and
the small circles with ${\bf S}_\mu(0) \parallel {\bf c}$ configuration at PSI. 
Both measurements were carried out using different batches of crystals. 
$\chi$ was measured in both zero-field-cooling $ZFC$ and field-cooling $FC$ mode 
with $H=1$~kOe. 
The inset of (b) shows the $T$ dependence of 
the slope of $\chi$ (d$\chi$/d$T$) 
measured in $ZFC$. 
}
\end{figure}
Figures~\ref{fig:ZF-muSR}(a) and \ref{fig:ZF-muSR}(b) show 
the $T$ dependence of  
the muon precession frequencies ($\nu_{\mu, i}$=$\omega_{\mu, i}/2\pi$) 
and $\chi$ 
for the single crystal platelets of K$_{0.49}$CoO$_2$. 
According to the $T$ dependence of the FFT, 
as $T$ increases from 1.63~K, 
$\nu_{\mu, 2}$ is roughly $T$ independent, 
while $\nu_{\mu, 1}$ and $\nu_{\mu, 3}$ decrease monotonically
with increasing $T$ up to $\sim$16~K. 
Then, the lowest $\nu_{\mu, 3}$ disappears, 
while $\nu_{\mu, 1}$ and $\nu_{\mu, 2}$ suddenly decreases by about 0.3~MHz. 
With further increase of  $T$, 
the two frequencies decrease monotonically with increasing slope (d$\nu_{\mu}$/d$T$)
and disappear at $\sim$55~K. 
This behavior is quite consistent with the $\chi$(T) curve, 
which exhibits a small peak at 55~K and a rapid increase below 20~K. 
Also, the $\chi$(T) curve measured in $ZFC$ mode starts to deviate 
from that in $FC$ mode below 20~K.
We therefore conclude that K$_{0.49}$CoO$_2$ undergoes 
a transition from a paramagnetic to 
a magnetically ordered state at $T_{\rm C,1}$=55~K
and then to the other ordered state at $T_{\rm C,2}\sim$16~K. 


Although neutron diffraction experiments are the most powerful technique 
to determine the magnetic structures of the two phases, 
the current $\mu^+$SR results provide primary information on them. 
Firstly, there are no indications of charge-ordering 
in the $\rho$(T) and $\chi(T)$ curve for K$_{0.49}$CoO$_2$, 
in contrast to both Na$_{0.5}$CoO$_2$ and K$_{0.5}$CoO$_2$.\cite{KCO_2} 
This suggests that the commensurate, three-sublattice,
120$^{\rm o}$ twisted {\sf SDW} state is most unlikely to exist in K$_{0.49}$CoO$_2$, 
because the 120$^{\rm o}$ AF domain structure, 
which is the ground state for the classical AF triangular lattice, 
naturally induces an insulating state.  
Secondly, the isotropic behavior observed 
by the ZF-$\mu^+$SR measurements 
clearly precludes in this case the possibility of the $A$-type AF ---
i.e. FM order in the CoO$_2$ plane 
but AF between the adjacent two CoO$_2$ planes, 
which was proposed for Na$_{0.82}$CoO$_2$.\cite{MPI_1} 

\begin{figure}
\includegraphics[width=3.5cm]{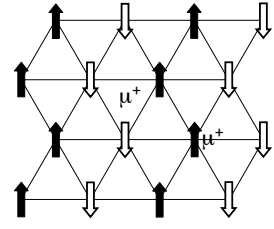}
\caption{\label{fig:LSDW}  Spin structure of a linear spin density wave 
({\sf LSDW}) state on the two-dimensional triangular lattice. 
Two possible muon sites ($\mu^+$), 
which locate not in the Co plane but in the oxygen plane and are surrounded 
by three O$^{2-}$ ions, are shown. 
These two sites are more preferable than the sites between the edge of the triangle,
which are surrounded by two O$^{2-}$ ions.
}
\end{figure}

\begin{figure}
\includegraphics[width=6.5cm]{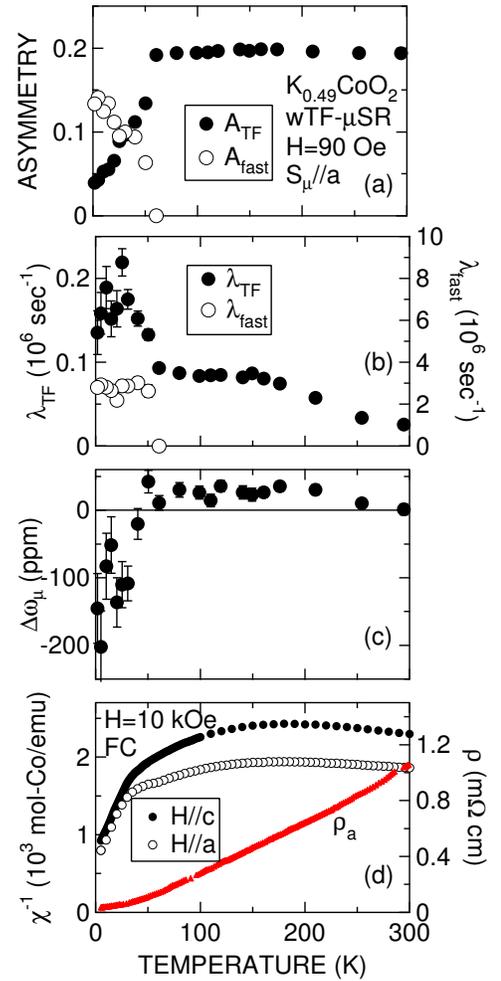}
\caption{\label{fig:wTF}  Temperature dependences of 
(a) $A_{\sf TF}$ and $A_{\sf fast}$, 
(b) $\lambda_{\sf TF}$ and $\lambda_{\sf fast}$ and 
(c) the shift of the muon precession frequency, $\Delta\omega_{\mu}$ and 
(e) the inverse susceptibility, $\chi^{-1}$ and resistivity, $\rho$\cite{KCO_1} 
in K$_{0.49}$CoO$_2$ crystals. 
The data were obtained by fitting 
the wTF-$\mu^+$SR spectra with 
a combination of a slowly relaxing precessing signal due to the external field and 
a fast non-oscillatory signal caused by the internal fields; 
$A_{\rm TF}\exp(-\lambda_{\rm TF})\cos(\omega_{\mu}t~+~\phi)~
+~A_{\rm fast}\exp(-\lambda_{\rm fast}t)$.
}
\end{figure}
In order to explain magnetism on the two-dimensional triangular lattice, 
the Hubbard model within a mean field approximation has been applied 
using the Hamiltonian ${\cal H}$ given by
\cite{2DTL_1,2DTL_2}
\begin{eqnarray}
 {\cal H}&=&-t\sum_{<ij>\sigma}c_{i\sigma}^{\dagger}c_{j\sigma} + 
 U\sum_i n_{i\uparrow}n_{i\downarrow} ,
\label{eq:Hubbard}
\end{eqnarray}
where $c_{i\sigma}^{\dagger}(c_{j\sigma})$ 
creates (destroys) an electron with spin $\sigma$ on site $i$, 
$n_{i\sigma}=c_{i\sigma}^{\dagger}c_{i\sigma}$ 
is the number operator, 
$t$ is the nearest-neighbor hopping amplitude and 
$U$ is the Hubbard on-site repulsion.
The electron filling $n$ is defined as 
$n$=(1/2$N$)$\sum_i^N n_i$,
where $N$ is the total number of sites.  

Since decrease in the K content ($x$) produces holes in K$_{x}$CoO$_2$, 
$x$=0.4 corresponds to $n$=0.3, 
if we ignore oxygen deficiency in the sample. 
As $n$ decreases from 0.5, 
the calculations predict that the linear ({\sf L}) {\sf SDW} phase 
transforms into the helical ({\sf H}) {\sf SDW} phase
at around $n$=0.4 for 4.4$\leq U/t\leq$4.8. 
For the {\sf LSDW} phase, a zigzag FM chain is formed 
and neighboring FM chains align antiferromagnetically (see Fig.~$\ref{fig:LSDW}$),\cite{LSDW_1} 
while Co spins form a helical structure in the plane
for the {\sf HSDW} phase.

In the {\sf LSDW} state, as seen in Fig.~$\ref{fig:LSDW}$, 
two possible muon sites, 
one just above/below the corner of the triangular lattice  
and the other at the center of the triangular lattice, 
originate the two different $\nu_{\mu}$s, 
corresponding to the two sharp peaks in the FFT above 20~K. 
In the {\sf HSDW} state, however, the helical spin structure 
produces a wide field distribution at the muon sites. 
In other words, the {\sf LSDW} state is most likely to appear below 55~K, 
and then the {\sf HSDW} state below $\sim$16~K. 

In order to elucidate the magnetic behavior above 55~K, 
we performed weak-transverse-field (wTF-) $\mu^+$SR 
measurements up to 300~K, 
with the results shown in Fig.~$\ref{fig:wTF}$ together with $\chi^{-1}$ and $\rho$. 
Besides the two transitions at 55~K and 20~K, 
both the slope of the wTF-relaxation rate ($\lambda_{\rm TF}$) and 
the normalized wTF-oscillation frequency ($\Delta\omega_{\mu}$) 
show an anomaly at $\sim$150~K, 
where the $\chi^{-1}(T)$ curve exhibits a broad maximum. 
Above 60~K, the wTF-asymmetry ($A_{\rm TF}$) levels off to
its maximum value ($\sim$0.2)
--- i.e. the sample volume is almost 100\% paramagnetic.  
This therefore suggests that 
the changes in $\lambda_{\rm TF}$ and $\Delta\omega_{\mu}$ are due to  
a spin state transition.  
The other possibility, a charge-ordering of Co$^{3+}$ and Co$^{4+}$,  
is most unlikely to occur, 
because the $\rho(T)$ curve shows no anomalies around 150~K. 
A spin state transition was also proposed for PrCoO$_3$,\cite{LnCO_1} 
for which the $\chi^{-1}(T)$ curve exhibits a broad maximum around 200~K. 

The wTF-$\mu^+$SR results also provide an insight for the ordered state.
The monotonic decrease in the $A_{\rm TF}$ below 55~K 
(and the accompanying increase in $A_{\rm fast}$) indicates 
that the magnetic order develops gradually with decreasing $T$ 
due to geometrical frustration of the triangular lattice. 
Actually, the $A_{\rm TF}(T)$ curve does not level off at the lowest $T$ measured.  
The volume fraction of the magnetic phase extrapolated to 0~K is estimated as $\sim$80$\%$. 

It should be noted that the overall $T$ dependence of $\nu_{\mu,i}$ 
for K$_{0.49}$CoO$_2$ is very similar to 
that for the Na$_{0.5}$CoO$_2$ sample.\cite{Mendels_1} 
Since in that experiment the sample was
polycrystalline and the data reported only below 90~K, there was
no information on possible magnetic anisotropy
or a spin state transition. 
This therefore raises the question concerning 
the origin of the magnetic order in Na$_{0.5}$CoO$_2$. 
Although alkali ions become more ionic with increasing atomic number, 
there are almost no contributions to the shape of the electronic structure, 
but just producing holes, even for NCO.\cite{singh_1} 
The main difference between NCO and K$_{0.49}$CoO$_2$ 
is hence the change in the interlayer distance of the two adjacent CoO$_2$ planes 
and as a result the decrease in the interlayer interaction. 

The present results indicate that 
magnetism in the layered cobaltites with $V_{\rm Co}\sim$3.5 
is not strongly affected by interlayer distance/interaction 
and the charge-ordering 
but by $V_{\rm Co}$. 
In other words, 
the phase diagram proposed for NCO\cite{Foo_1} 
is likely to be applicable to the other $A_x$CoO$_2$, 
if the $x$ region of the magnetic order is extended.  
Additionally, we have performed $\mu^+$SR measurements on Rb$_{0.3}$CoO$_2$, 
which showed the absence of magnetism down to 5~K. 
The ZF-$\mu^+$SR spectra in that case are well described  by a dominant
static Kubo-Toyabe function 
with the field distribution $\Delta\sim$0.2$\times$10$^6$~s$^{-1}$ at 5~K, 
implying that the muons experience  interactions only with the spins of 
Co and Rb nuclei in a 
paramagnetic state, 
similar to Na$_{0.35}$CoO$_2$.\cite{N035CO-NMR_1} 
This further supports the similarity among the $A_x$CoO$_2$ family. 
We wish finally to mention that, 
if the K/$A$ content is controlled precisely, 
better thermoelectrics and even new superconductors 
are also likely to be found in $A_x$CoO$_2$. 
  
This work was performed at both TRIUMF and PSI. 
We thank S.R. Kreitzman, B. Hitti, D.J. Arseneau of TRIUMF, 
P. Russo of Columbia Univ. and the 
LMU staff of PSI 
for help with the $\mu^+$SR experiments, 
and 
Mr. Y. Kawai of Toyota CRDL for the ICP analysis. 
This work was partially (JHB) supported 
at UBC by CIAR, NSERC of Canada, 
and at TRIUMF by NRC of Canada.


\end{document}